\documentclass[showpacs,amsmath,amssymb,twocolumn]{revtex4}
\usepackage{epsfig}
\input{epsf}
\usepackage{amssymb}
\begin{document}
\title{Phantom universe from CPT symmetric QFT}

\author{Alexander A. Andrianov}
\affiliation{Dipartimento di Fisica and INFN, Via Irnerio 46,40126 Bologna,
Italy\\
V.A. Fock Department of Theoretical Physics, Saint Petersburg State University,
198904, S.-Petersburg, Russia} 
\author{Francesco Cannata}
\affiliation{Dipartimento di Fisica and INFN, Via Irnerio 46,40126 Bologna,
Italy}
\author{Alexander Y. Kamenshchik}
\affiliation{Dipartimento di Fisica and INFN, Via Irnerio 46,40126 Bologna,
Italy\\
L.D. Landau Institute for Theoretical Physics of the Russian
Academy of Sciences, Kosygin str. 2, 119334 Moscow, Russia}

\begin{abstract} 
Inspired by the generalization of quantum theory for the case of non-Hermitian
Hamiltonians with CPT symmetry, we construct a simple classical cosmological
scalar field based 
model describing a smooth transition from ordinary dark energy to the 
phantom one. 
\end{abstract}
 \maketitle 

The discovery of the cosmic acceleration \cite{cosmic} has stimulated 
the search for models of the so called dark energy \cite{dark} responsible
for this phenomenon. The crucial feature of the dark energy is that 
$w=p/\varepsilon < -1/3$, where $p$ is the pressure and $\varepsilon$ 
is the energy density. Some observations \cite{observ} hint to the possibility
that the equation of state parameter $w < -1$. The corresponding models 
are called phantom dark energy ones \cite{phantom}. These models have 
some unusual properties: to realize them one often uses the phantom scalar 
field with the negative sign of kinetic term; in many models the presence 
of the phantom dark energy implies the existence of the future Big Rip 
cosmological singularity \cite{Rip}; according to some observations 
the crossing of the phantom divide line $w = -1$ occurs, the theoretical 
explanation of this fact also presents some kind of challenge \cite{cross}.

The phantom model building has involved many different ideas. Here we would 
like to present a rather simple and natural cosmological toy model, 
linked to and inspired by such an intensively developing branch of 
quantum mechanics and 
quantum field theory as the study of non-Hermitian, but $CPT$ (or $PT$) 
symmetric models \cite{BB,non-Hermit,QFT,non-Hermit1}.          
The main point of this approach consists in the fact that there exists 
a large class of non-Hermitial Hamiltonians, which nevertheless possesses
 real and often positive definite spectrum. 
As found by Bender and Boettcher  in their seminal paper \cite{BB},
they are characterized by 
a potential which in one-dimensional case satisfies the property of $PT$ -
invariance $V(x) = V^*(-x)$. Non-trivial generalization to quantum field 
theory has also been considered \cite{QFT}. It has been suggested that  
non-Hermitian quantum theory may find applications in quantum cosmology
\cite{non-Hermit1}. 

Here, we explore the use of a particular  complex scalar field 
Lagrangian, which has real solutions of the classical equations of motion.
Thereby we provide a cosmological model describing 
in a natural way an 
evolution from the Big Bang to the Big Rip involving the transition from 
normal matter to phantom matter, crossing smoothly the phantom divide line. 
The interest of our approach is related to its focusing on the 
intersection between two important fields of research, hopefully 
allowing for a their mutual cross-fertilization.

In particular, we give an example of charged scalar matter interacting with 
a non-Hermitian potential which however does not break
the CPT symmetry. In our model the classical solutions 
in the presence of gravity ( FRW cosmological background)
are such that the originally complex Lagrangian becomes real on classical 
vacuum configurations while one of the scalar component
obtains the ghost sign of kinetic energy. Thereby we recover a more 
conventional phantom matter starting from the complex
matter with normal kinetic energy.

Let us consider a CPT symmetric, but non-Hermitian Lagrangian of a
scalar field
\begin{equation}
L = \frac12\partial_{\mu}\phi\partial^{\mu}\phi^* - V(\phi,\phi^*),
\label{Lagrange}
\end{equation}
with a potential $V(\phi,\phi^*)$ satisfying the CPT invariance condition
\begin{equation}
(V(\phi,\phi^*))^* = V(\phi^*,\phi),
\label{condition}
\end{equation}
while the condition 
\begin{equation}
(V(\phi,\phi^*))^* = V(\phi,\phi^*),
\label{condition1}
\end{equation}
is not satisfied. (Indeed, it should be noted that P-invariance is trivial
for a scalar field).
For example, such potential can have a form
\begin{equation}
V(\phi,\phi^*) = V_1(\phi+\phi^*)V_2(\phi-\phi^*).
\label{factor}
\end{equation}
If one defines 
\begin{equation}
\phi = \phi_1 + i\phi_2
\label{split}
\end{equation}
and considers potentials of the form
\begin{equation}
V(\phi,\phi^*) = V_0(\phi_1)\exp(i\alpha\phi_2),
\label{factor1}
\end{equation}
where $\alpha$ is real parameter a, one can recognize the link to 
the so called $PT$ symmetric potentials.

Here, the functions $\phi_1$ and $\phi_2$ are introduced as 
the real and the imaginary parts of the complex scalar field $\phi$,
however, in what follows, we shall treat them as independent spatially 
homogeneous 
variables depending only on the time parameter $t$.
The equations of motion for fields $\phi_1$ and $\phi_2$ have the form
\begin{equation}
\ddot{\phi_1} + 3h\dot{\phi_1} +V_0'(\phi_1)\exp(i\alpha\phi_2) = 0,
\label{motion}
\end{equation}
\begin{equation}
i\ddot{\phi_2}+ 3ih\dot{\phi_2} - \alpha V_0(\phi_1)\exp(i\alpha\phi_2) = 0,
\label{motion1}
\end{equation}
where $h \equiv \frac{\dot{a}}{a}$ is the Hubble variable 
for a flat spatially homegeneous metric
\begin{equation}
ds^2 = dt^2 -a^2(t)dl^2,
\label{Fried}
\end{equation}
satisfying the Friedmann equation
\begin{equation}
h^2 = \frac12\dot{\phi}_1^2 + \frac12\dot{\phi}_2^2 + 
V_0(\phi_1)\exp(i\alpha\phi_2).
\label{Fried1}
\end{equation} 
Let us notice, that the system of equations 
(\ref{motion}),(\ref{motion1}),(\ref{Fried1}) can have a solution 
where $\phi_1(t)$ is real, while the $\phi_2$ is imaginary, or, in other words
\begin{equation}
\phi_2(t) = -i\xi(t), 
\label{imaginary}
\end{equation}
where $\xi(t)$ is a real function. In terms of these 
two real functions, our system of equations can be rewritten as
\begin{equation}
\ddot{\phi_1} + 3\sqrt{\frac12\dot{\phi}_1^2 - \frac12\dot{\xi}^2 + 
V_0(\phi_1)\exp(\alpha\xi)}
\dot{\phi_1} +V_0'(\phi_1)\exp(\alpha\xi) = 0,
\label{motion2}
\end{equation}
\begin{equation}
\ddot{\xi} + 3\sqrt{\frac12\dot{\phi}_1^2 - \frac12\dot{\xi}^2 + 
V_0(\phi_1)\exp(\alpha\xi)}
\dot{\xi} -\alpha V_0(\phi_1)\exp(\alpha\xi) = 0.
\label{motion3}
\end{equation}

Now, substituting $\phi_2(t)$  from Eq. (\ref{imaginary}) into Eq. 
(\ref{Fried1}) we have the following expression for the energy density
\begin{equation}
\varepsilon = h^2 = \frac12\dot{\phi}_1^2 - \frac12\dot{\xi}^2 + 
V_0(\phi_1)\exp(\alpha\xi).
\label{energy}
\end{equation} 
The pressure will be equal 
\begin{equation}
p = \frac12\dot{\phi}_1^2 - \frac12\dot{\xi}^2 - 
V_0(\phi_1)\exp(\alpha\xi).
\label{pressure}
\end{equation} 
It is  easy to see that if $\dot{\phi}_1^2 < \xi^2$ the pressure will be 
negative and $p/\varepsilon < -1$,  satisfying the phantom equation 
of state. Instead, when $\dot{\phi}_1^2 > \xi^2$, the ratio between 
the pressure and energy density exceeds $-1$ and, hence, the 
condition 
\begin{equation}
\dot{\phi}_1^2 = \dot{\xi}^2 
\label{divide}
\end{equation}
corresponds  exactly to the phantom divide line, which can be crossed 
dynamically during the evolution of the field components $\phi_1(t)$ and 
$\xi(t)$.  

We provide  now a simple realization of this idea by
an exactly solvable cosmological model by implementing the
technique for construction of potentials for a given 
cosmological evolution \cite{evolution}. 
It is convenient to start with a cosmological evolution as given 
by the following expression for the Hubble variable: 
\begin{equation}
h(t) = \frac{A}{t(t_R-t)}.
\label{evolution}
\end{equation}
The evolution begins at  $t =0$, which represents a standard initial Big Bang  
cosmological singularity, and comes to an end in the Big Rip type 
singularity at $t = t_R$.  The derivative of the Hubble variable 
\begin{equation}
\dot{h} = \frac{A(2t - t_R)}{t^2 (t_R -t)^2}
\label{derivative}
\end{equation}
vanishes at  
\begin{equation}
t_P = \frac{t_R}{2}
\label{PDL}
\end{equation}
when the universe crosses the phantom divide line.

Next, we can write down the standard formulae connecting the energy density 
and the pressure to the Hubble variable and its time derivative:
\begin{equation}
\frac{\dot{\phi}_1^2}{2} - \frac{\dot{\xi}^2}{2} + V_0(\phi_1)e^{\alpha\xi} = 
h^2 = \frac{A^2}{t^2(t_R-t)^2},
\label{Hubble-en}
\end{equation}
\begin{equation}
\frac{\dot{\phi}_1^2}{2} - \frac{\dot{\xi}^2}{2} - V_0(\phi_1)e^{\alpha\xi} = 
-\frac23\dot{h}-
h^2 = -\frac{A(4t-2t_R+3A)}{3t^2(t_R-t)^2}.
\label{Hubble-pres}
\end{equation}
The expression for the potential $V_0(\phi_1)$ follows 
\begin{equation}
V_0(\phi_1) = \frac{A(2t-t_R+3A)}{3t^2(t_R-t)^2}e^{-\alpha\xi}.
\label{potential}
\end{equation}
The kinetic term satisfies the equation
\begin{equation}
\dot{\phi}_1^2 - \dot{\xi}^2 = -\frac{2A(2t-t_R)}{3t^2(t_R-t)^2}.
\label{kinetic}
\end{equation}

It is convenient to begin the construction with
the solution for $\xi$. 
Taking into account the formulae (\ref{evolution}) and (\ref{potential}) 
Eq. (\ref{motion3}) can be rewritten as 
\begin{equation}
\ddot{\xi} + 3\dot{\xi} \frac{A}{t(t_R-t)}  
-\frac{\alpha A(2t-t_R+3A)}{3t^2(t_R-t)^2} = 0.
\label{motionxi}
\end{equation}
Introducing a new parameter
\begin{equation}
m \equiv \frac{3A}{t_R},
\label{mdefine}
\end{equation}
Eq. (\ref{motionxi}) looks like 
\begin{equation}
\dot{y} + y \frac{mt_R}{t(t_R-t)} - 
\frac{\alpha mt_R(2t+t_R(m-1))}{9t^2(t_R-t)^2} = 0,
\label{motionxi1}
\end{equation}
where 
\begin{equation}
y \equiv \dot{\xi}.
\label{ydefine}
\end{equation}
It is not difficult to show that the solution of Eq. (\ref{motionxi1}) 
 is given by
\begin{equation}
y = \frac{\alpha mt_R(t_R-t)^m}{9t^m} \int dt \frac{(2t + (m-1)t_R)t^{m-2}}
{(t_R-t)^{m+2}},
\label{integral}
\end{equation} 
where the inessential constant of integration will be disregarded.

Let us estimate the behavior of the solution (\ref{integral}) at 
$t \rightarrow t_R$. Simple estimation gives
\begin{equation}
y \rightarrow -\frac{\alpha m}{9(t_R-t)}.
\label{estimation}
\end{equation}
On the other hand the equality (\ref{kinetic}) should be satisfied for 
all the values $t \leq t_R$. That means that the value of $\dot{\xi}^2$ 
should be grater than the absolute value of right-hand side of 
Eq. (\ref{kinetic}). This last quantity at the limit $t \rightarrow t_R$ 
behaves as $2m/9(t_R-t)^2$.
Thus, one should have 
\begin{equation}
\frac{\alpha^2 m^2}{81(t_R-t)^2} \geq \frac{2m}{9(t_R-t)^2}
\label{ineq}
\end{equation} 
or, in other words,
\begin{equation}
m \geq \frac{18}{\alpha^2}.
\label{ineq1}
\end{equation}
Before considering the concrete values of $m$, notice 
that the equation of state parameter $w$ in the vicinity 
of the initial Big Bang singularity behaves as 
\begin{equation}
w = -1 + \frac{2}{m},
\label{BigBang}
\end{equation}
while approaching the final Big Rip singularity this parameter 
behaves as
\begin{equation}
w = -1 - \frac{2}{m}.
\label{BigRip}
\end{equation}
Notice that the range for $w$ does not depend on $\alpha$, depending 
only on the value of the parameter $m$, which relates the scales of the 
Hubble variable $h$ and of the time of existence of the universe $t_R$.

Remarkably, an integral in the right-hand side of Eq. (\ref{integral}) 
is calculable analytically 
\begin{equation}
\dot{\xi} = \frac{\alpha mt_R}{9t(t_R-t)}
\label{xiint}
\end{equation}
while 
\begin{equation}
\xi = \frac{\alpha m}{9} (\log t - \log(t_R-t)).
\label{xiint1}
\end{equation}
From now on the parameter $t$ will be dimensionless. Inclusion of
  characteristic time does not change the structure of the potential because 
of its exponential dependence on $\xi$.
Substituting the expression (\ref{xiint}) into Eq. (\ref{kinetic}) one 
has
\begin{equation}
\dot{\phi}_1^2 = \frac{mt_R((\alpha^2 m+18)t_R - 36t)}{81t^2(t_R-t)^2}.
\label{phieq}
\end{equation}
For the case $\alpha^2 m = 18$ the function $\phi_1(t)$ can be easily found 
from Eq. (\ref{phieq}) and it looks like follows:
\begin{equation}
\phi_1 = \pm \sqrt{32} {\rm Arctanh} \sqrt{\frac{t_R-t}{t_R}}.
\label{phitime}
\end{equation}
One can choose the positive  sign in Eq. (\ref{phitime}) without 
loosing the generality. 

Inverting Eq. (\ref{phitime}) we obtain the dependence of the time 
parameter as a function of $\phi_1$
\begin{equation}
t = \frac{t_R}{\cosh^2\frac{\phi_1}{\sqrt{32}}}.
\label{timephi}
\end{equation} 
Substituting expressions (\ref{timephi}) and 
(\ref{xiint1}) into Eq. (\ref{potential}) we can obtain 
the explicit expression for the potential $V_0(\phi_1)$:
\begin{equation}
V_0(\phi_1) = \frac{2\cosh^6\frac{\phi_1}{\sqrt{32}}\left(2 + 17
\cosh^2\frac{\phi_1}{\sqrt{32}}\right)}{t_R^2}.
\label{potential-result}
\end{equation}
We would like to emphasize that this potential is real and even.
It is interesting that the time dependence of $\phi_1(t)$ could be 
found also for an arbitrary value of the parameter $m$, but for 
$\alpha^2 m > 18$  this dependence cannot be reversed analytically and, 
hence, one cannot
obtain the explicit form of the potential $V_0(\phi_1)$. 

Now, let us turn to Eq. (\ref{timephi}), expressing the dependence 
of the time parameter $t$ on $\phi_1$. It is convenient to consider 
the evolution of the value of $\phi_1$ between the values $-\infty$, 
corresponding to the initial cosmological singularity at $t = 0$,  
and $0$, corresponding to the Big Rip singularity at $t = t_R$. 
At the moment $t_P = t_R/2$, the equality $\dot{\xi}^2 = \dot{\phi}_1^2$
is satisfied and the universe is crossing the phantom divide line.
It is easy to obtain from Eqs. (\ref{xiint1}) and (\ref{phitime}) the 
 values of the fields $\xi$ and $\phi_1$ at $t=t_P$:
\begin{equation}
\xi(t_P) = 0,\ \ \phi_1(t_P) = \sqrt{32}{\rm Arctanh} \sqrt{\frac12}.
\label{values}
\end{equation} 
The potential (\ref{potential-result}) is smooth together with all its
derivatives at this point. Moreover, it is quite regular at all 
the finite values of the scalar field $\phi_1$. It is curious that 
the potential $V_0(\phi_1)$ is finite also at the moment of the Big Rip.
It is not, however, strange, because it enters in the expressions for the 
energy density (\ref{Hubble-en}) and the pressure (\ref{Hubble-pres})
being multiplied by the factor $e^{\alpha\xi}$ which is singular at $t = t_R$. 

In conclusion we would like to stress that we have constructed a model
 relaxing the requirement of Hermiticity of the Hamiltonian of
the theory which is equivalent to the reality of the classical Lagrangian.
This relaxation, however, does not imply the breakdown of Lorentz and 
CPT invariance. For 
our classical solutions, expressed in terms of real fields,  observable 
quantities like energy density, pressure, Hubble variable turn out to be real. 
As a consequence our model describes in a rather natural way the transition
from  normal matter to  phantom one.

\section*{Acknowledgments}
A.A. was partially supported by RFBR, grant No. 05-02-17477 and 
by the Programme UR, grant No. 02.01.299. 
A.K. was  partially supported by  RFBR, grant No. 05-02-17450 and  
by the Research Programme "Astronomy" of the RAS.

\end{document}